\documentclass[]{aastex701}
\usepackage{amsmath,amssymb,amsfonts}%
\definecolor{commentt}{RGB}{128,128,96}
\definecolor{rkka}{RGB}{219,66,32}
\definecolor{gree}{RGB}{19,96,22}
\newcommand{\eb}{\begin{equation}}
\newcommand{\ee}{\end{equation}}
\shorttitle{Space Velocity of Messier 87}
\shortauthors{Makarov et al.}
\begin{document}
\title{The Peculiar Velocity of Messier~87 from Microarcsecond Geodetic VLBI Astrometry}
\correspondingauthor{Valeri Makarov}
\author[orcid=0000-0003-2336-7887,gname=Valeri,sname=Makarov]{Valeri V. Makarov}
\affiliation{U.S. Naval Observatory, 3450 Massachusetts Ave NW, Washington, DC 20392-5420, USA}
\email[show]{valeri.v.makarov.civ@us.navy.mil}  
\author[orcid=0000-0002-8736-2463,gname=Phil,sname=Cigan]{Phil Cigan}
\affiliation{U.S. Naval Observatory, 3450 Massachusetts Ave NW, Washington, DC 20392-5420, USA}
\email{phillip.j.cigan.civ@us.navy.mil}
\author[orcid=0000-0002-2764-7248,gname=Kyle,sname=Corcoran]{Kyle A. Corcoran}
\affiliation{U.S. Naval Observatory, 3450 Massachusetts Ave NW, Washington, DC 20392-5420, USA}
\affiliation{Computational Physics, Inc., 8001 Braddock Rd. Suite 210, Springfield, 22151-2110, VA, USA}
\email{kyle.a.corcoran3.ctr@us.navy.mil}

\begin{abstract}   
Our knowledge of the space velocity of Messier 87, which is the dominant galaxy in the Virgo cluster, has been limited to the radial velocity component. Using a cadence of precision position measurements with the global geodetic very long baseline interferometry (VLBI) system over 28 years, we determined the proper motion vector of the radio-emitting core by a robust statistical method involving 1-norm optimization and bootstrapping. The proper motion vector is directed at a position angle $189.2\degr \pm 3.5\degr$ in the equatorial International Celestial Reference Frame, and its magnitude is $10.19$ $\mu$as yr$^{-1}$ with an uncertainty of $0.64$ $\mu$as yr$^{-1}$. The projected velocity of the AGN in the tangential sky plane is ($787\pm50$)~km~s$^{-1}$. The peculiar velocity of Messier 87 with respect to the preferred rest frame of the cosmic microwave background field is approximately 1037 km s$^{-1}$ (assuming a distance of 16.1 Mpc) with an angle of 65$^\circ$ to the current line of sight, which implies a tangential relative motion of M87 and the Galaxy. The peculiar velocity of M87 is directionally concordant with the reconstructed and $\Lambda$CDM-simulated motion of the Virgo filament towards the Great Attractor, but the Milky Way moves slower by 470 ~km~s$^{-1}$ in that direction.
\end{abstract}



\section{Introduction}\label{sec1}

Messier 87, hereafter M87, is the central elliptical galaxy in the Virgo cluster, situated near its center of mass \citep{1994Natur.368..828B}. The Virgo cluster is a crucially important rung in the cosmic distance ladder \citep{1994Natur.371..385P, 2013IAUS..289..262M}. The entire edifice of the $\Lambda$CDM model is based on the local determination of the Hubble constant $H_0$, which requires calibration of primary and secondary distance scale tracers at the highest distances where they can still be resolved. 
Our Galaxy belongs to the Virgo Supercluster, the largest kind of matter agglomeration in the Universe allowed by the concordance $\Lambda$CDM model, where the Virgo cluster is the central substructure. Knowledge of three-dimensional velocities and accelerations of different parts of the local superstructure with respect to the Galaxy is essential for testing against the predictions of $\Lambda$CDM, which set constraints on deviations of intercluster peculiar velocities from the Hubble flow. 
On a smaller scale, the kinematic structure is important to correctly map the local configuration and dynamical history of the Local Supercluster (LSC) \citep{1984ApJ...281...31T}. 

In cosmologies such as $\Lambda$CDM based on the Friedman-Lema{\^i}tre-Robertson-Walker metric, the evolution of the Universe is governed by a single scale parameter $a(t)$, allowing isotropic expansion in which galaxy clusters and quasars appear to recede from the comoving observer in a purely radial direction that is invariant to the observer's location, i.e., there is no systemic tangential motion due to the cosmic expansion. In alternative anisotropic spacetimes, as well as in the radially inhomogeneous Lema{\^i}tre-Tolman-Bondi metrics, the cosmic expansion rate becomes not only time-dependent but also direction- and distance-dependent because of the offset of the observer from the center of the Universe (i.e., the geometric center of cosmic expansion). This generates a systemic pattern of tangential velocities, which is sometimes called the ``cosmic parallax"\footnote{``Parallax" is perhaps a misnomer in the context, as this term is traditionally reserved for an effect caused by the motion of the observer.} in the literature \citep{2009PhRvL.102o1302Q}. Detecting such a pattern with precision astrometry of distant quasars would be revolutionary for our understanding of the Universe. The main obstacles for this quest are the limited signal-to-noise ratio of present-day astrometric measurements, a limited number of available sources, and the actual peculiar velocities of quasars and AGNs. For a long time believed to be negligible, these individual deviations from the general cosmic flow begin to emerge as a considerable factor in light of recent advances in astrometry. Whether the observed motions of distant quasars are statistically independent and random is yet to be seen.

The methods of optical astrometry have seen a remarkable advancement in the past decade owing to the successful and ambitious Gaia space mission  \citep{2016A&A...595A...1G}. However, even with the latest and most refined Data Release 3 \citep{2021A&A...649A...1G}, the level of astrometric precision is not sufficient to confidently determine the transverse angular velocities of galaxies beyond the Local Group. Those measured for the satellite galaxies of the Milky Way are dispersed within an upper bound of $\sim 400$ km s$^{-1}$ \citep{2020RNAAS...4..229M}, which probably reflects their orbital motion in the common gravitational potential. The median precision of these Gaia-based determinations is 50 $\mu$as yr$^{-1}$. A five-times better precision was achieved for the Andromeda galaxy (M31) \citep{2021MNRAS.507.2592S} owing to a much greater number of resolved member stars, confirming the previously deduced low tangential velocity of this dominant Local Group galaxy with respect to the center of the Milky Way. We will show in this work that the available time series of VLBI astrometric measurements allow for exploration of kinematics of more distant objects beyond the Local Group at a much higher level of precision well beyond 1 $\mu$as yr$^{-1}$.

Distance estimates for M87 range from  $15.4\pm 0.6$ Mpc \citep{2019AstBu..74..257T} to $17.2\pm 0.6$ Mpc \citep{2007ApJ...655..144M}.
The average distance over multiple determinations with various methods published in the literature is 16.07 Mpc with a sample standard deviation of 1.03 Mpc \citep{2020ApJS..246....3D}. Previous determinations of the apparent proper motion have been done on shorter time scales using various selections of VLBI global solutions in the S/X band \citep{2011yCat..35290091T, 2013yCat..35590095T, 2017ApJS..233....3T}. The latter study using 1740 individual measurements over 27.1 yr provides the highest formal precision and determines $\mu=\{-1.16,-8.74\}\pm \{0.94,1.10\}$ $\mu$as yr$^{-1}$ in the equatorial celestial coordinate system. Here, we use a special dedicated global solution to determine astrometric coordinates of M87 with a larger number of data points from astrometric and geodetic sessions, including five additional years of a multi-decadal observational campaign, to re-determine the proper motion magnitude and direction. Although the effect is strong in the data, a robust bootstrapping method is employed to realistically estimate the uncertainties of the derived parameters. The estimated tangential components of motion are combined with an independently estimated radial velocity and converted to a 3D velocity vector with respect to the Solar system barycenter. Taking into account the independently determined velocity of the barycenter relative to the cosmic microwave background (CMB) rest frame, we derive the `absolute' velocity and its direction in the CMB frame, as well as the relative motion of M87 with respect to the center of the Milky Way. The barycentric and absolute velocity vectors are transformed into the supergalactic coordinate system \citep{1976RC2...C......0D, 1998astro.ph..9343L} compared with the results of empirical reconstruction and constrained cosmological simulations in other publications.
\section{Data}

From a series of global solutions utilizing 6581 geodetic 24hr VLBI sessions from 1980 to 2023, astrometric positions of sources at each observing epoch were calculated as described in our earlier work \citep{2024ApJS..274...28C}.  
These time series data and global solution products are publicly hosted for download at the USNO web pages.\footnote{\url{https://crf.usno.navy.mil/quarterly-vlbi-solution}} 
The solutions used to create the source positions in this work are based on the extended usn2025a S/X band global solution, correcting for the Galactic aberration and including recent observations. Apart from the peculiar motion of nearby AGNs, the entire ensemble of distant sources in the Universe is involved in a coherent pattern of proper motions, which is independent of redshift. This relativistic effect is caused by the secular acceleration of the Solar system barycenter in the Galaxy as it travels in loose ``orbits" in the ambient gravitational potential \citep{2003A&A...404..743K, 2006AJ....131.1471K}. The acceleration vector is roughly directed toward the Galactic center. In the observer's frame, the apparent positions of all celestial sources are shifted in a dipole pattern from their ``inertial" directions by relatively large angles due to the aberration of light rays, but only the slow change of the aberrational dipole is astrometrically observable. 

To estimate the Galactic aberration for this special solution, we calculated the covariance-weighted mean value of the effect estimated from geodetic VLBI (V) observations and from Gaia (G) data. The emerging proper motion dipole can be represented with a vector $\boldsymbol{g}$ with Cartesian coordinates $(g_x,g_y,g_z)$, and the weighted mean vector is
\begin{equation}\label{eq: aberration mean}
\hat{\boldsymbol{g}} = \left(\boldsymbol{\Sigma}_\mathrm{V}^{-1} + \boldsymbol{\Sigma}_\mathrm{G}^{-1}\right)^{-1} \left(\boldsymbol{\Sigma}_\mathrm{V}^{-1} \boldsymbol{g}_\mathrm{V} + \boldsymbol{\Sigma}_\mathrm{G}^{-1} \boldsymbol{g}_\mathrm{G} \right),
\end{equation}
where $\boldsymbol{\Sigma}$ is the given formal covariance matrix.
\noindent Using the VLBI-determined dipole values from \cite{2013A&A...559A..95T} and the Gaia-determined values from \cite{2021A&A...649A...9G}, we find a proper motion dipole magnitude of $5.31\pm0.33$~$\mu$as~yr$^{-1}$ pointed towards $(\alpha, \delta) = (271.4\pm4.6, -30.9\pm3.6$) degrees, or $(l,b) = (0.5\pm3.6, -4.7\pm4.0$) degrees. For our geodetic analysis, we adopt a value of $5.3$~$\mu$as~yr$^{-1}$ and propagate the uncertainties on the resulting proper motion vectors of M87.

\begin{figure}[ht]%
\centering
\includegraphics[width=0.75\textwidth]{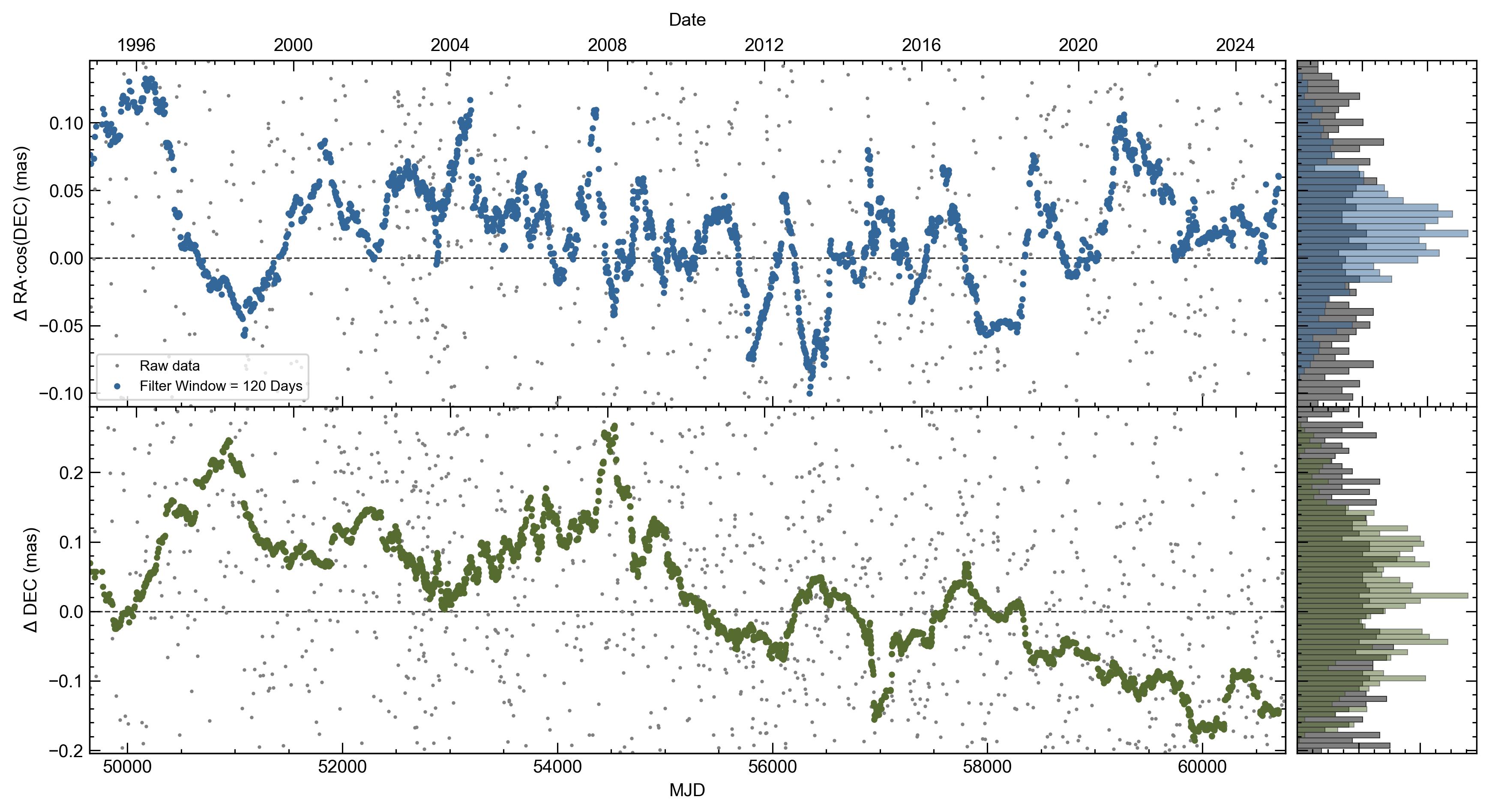}
\caption{Dots represent individual VLBI coordinate measurements obtained for M87 in right ascension (upper panel) and declination (lower panel) as functions of time. The sidewise histograms on the right are the sample distributions of these coordinate measurements. The colored circles are window-averaged values of the same measurements and highlight the underlying trends. Note that the scatter of individual data points before averaging is much greater than the vertical range of the window-averaged data, and the error bars are omitted to avoid graphical overcrowding.}\label{times.fig}
\end{figure}
 
\begin{figure}[h]%
\centering
\includegraphics[width=0.8\linewidth]{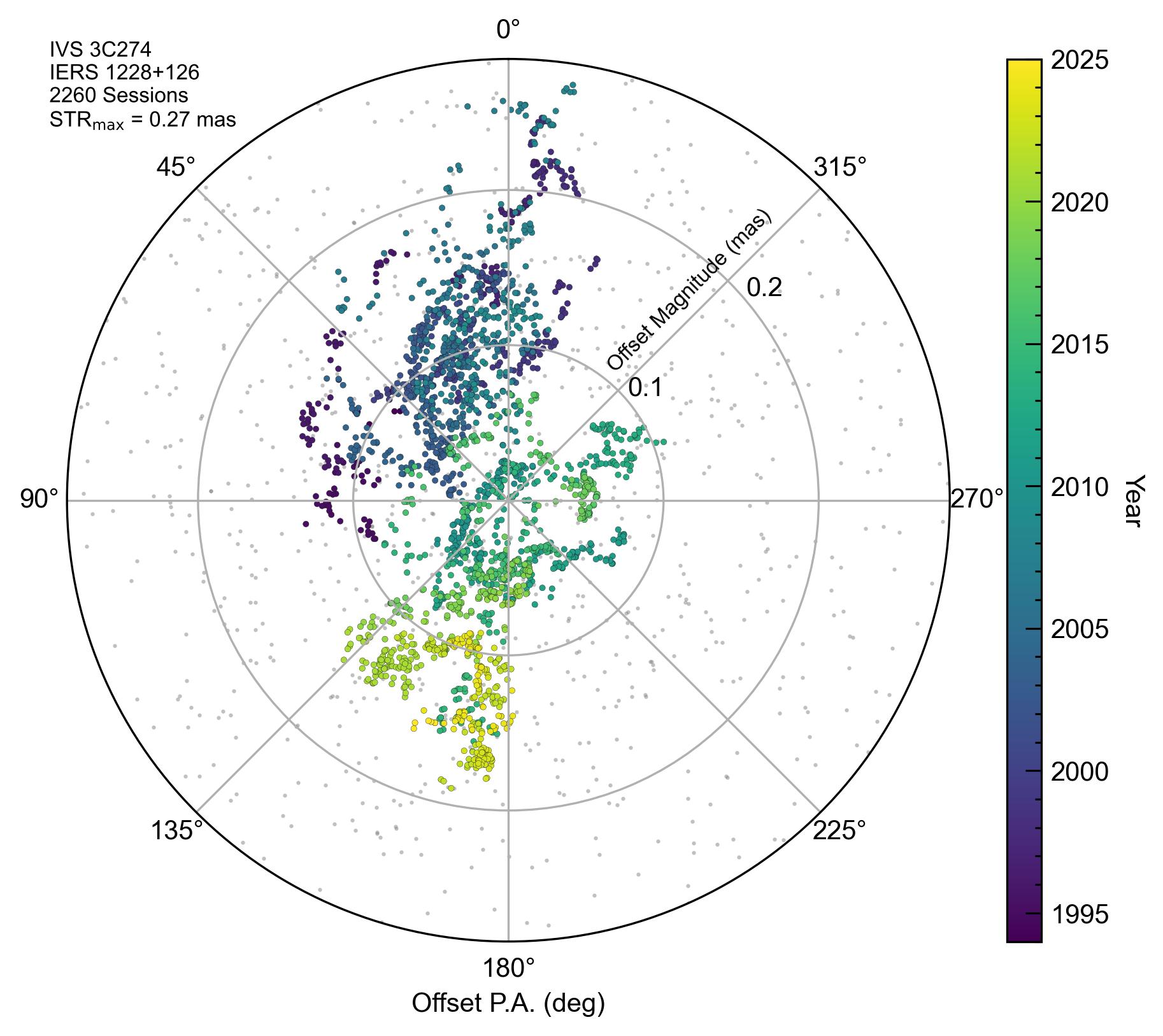}
\caption{VLBI time series positions on the sky for M87, smoothed with a 120-day window \citep{2024ApJS..274...28C}. Time is denoted by the color scale, while gray background points represent the unsmoothed time series.  This reveals clear coherent position variability on the order of $\sim$0.4 mas on several-year timescales.
This figure is available as an 8-second animation in the online version of this article.  In the animation, a scroll bar progresses smoothly through the years spanning the astrometric VLBI sessions, and the scatter points representing the single-epoch position estimates appear at the time of their observations, tracking the observed positions over time as they appear on the sky, ultimately ending with the static image where all observations are included. }
\label{VLBI_timeseries.fig}
\end{figure}

M87 was observed in 2269 geodetic S/X band sessions. 
The resulting time series consists of source right ascension and declination coordinates, with covariances of the position estimates at each epoch (once per day).  
Sessions prior to approximately MJD 49600 (~1996) were excluded from the analysis owing to significantly larger per-session uncertainties; their inclusion does not alter the derived proper motion.
The coordinate time series are displayed in Fig. \ref{times.fig}. Note that the scatter of individual data points (marked with gray dots) is much larger than the relevant range of these plots, and the corresponding side histograms are subsequently truncated. The colored circles mark the computed sliding-window averages over 4 month intervals, which help to reduce the measurement noise by a factor 10--11. The radical averaging allows us to see a complex pattern of short-term astrometric variations of yet unknown origin, and a general linear trend in the declination component. The distribution of window-averaged declination values is also clearly non-Gaussian.
A 2D overview of the VLBI position time series for M87 is displayed in Figure~\ref{VLBI_timeseries.fig}, with an animation of the time evolution and table of the data presented as supplements. Each point in this plot represents the sliding window-average of 120 days. This fixed time window averaging is more physically motivated and preferable to a fixed N-point technique because of the nonuniform observing cadence with irregular gaps in the schedule.

\section{Methods}
\label{meth.sec}

For initial analysis of the observed astrometric trajectory of M87, we employ a statistically robust non-parametric method to estimate the coherent component in the VLBI coordinate series, which is built on the concept of an astrometric differential autoregressive function (ADAF). This method is based on the intuitively clear observation that a long-term divergent trajectory in the sky plane will be seen as a generally increasing distance between individual positions separated by longer time intervals. For example, if the underlying ``systemic" part of the apparent trajectory is a diverging spiral, the ADAF in each coordinate looks like a wave with constantly increasing amplitude as a function of the time interval $\tau$ between two position measurements in each celestial coordinate. An object moving along a straight line with constant angular velocity yields an inclined straight ADAF in each sky coordinate. Compared to other validation and diagnostic methods, ADAF allows us to drastically reduce the measurement noise from $\sim 300$--$400$  to a few $\mu$as. It produces stable results for irregular cadences and explicitly non-Gaussian processes, and works well to reveal possible non-sinusoidal quasi-periodic and almost periodic signals.

The adopted implementation of the ADAF computation is as follows. Let $\{x_j,y_j\}$, $j=1,2,\ldots, n$ be the offsets of the celestial coordinates from the nominal mean position of a given source measured at times $t_j$ (the method is differential and, therefore, invariant to the choice of the nominal reference position). We construct the set of $n(n-1)/2$ unique pairs of data points with indices $\{k,m\}$, $m>k$. Each pair is separated by a time delay $\tau_{km} = t_m-t_k$, and the corresponding angular separation vector is $\boldsymbol{\delta}_{km}=[dx_{km},dy_{km}]^T=[x_m,y_m]^T-[x_k,y_k]^T$. 
The sequence of tuples $\{\tau_{km},\boldsymbol\delta_{km}\}$ is sorted by time delay values $\tau_{km}$. This sequence has a significant noise component because of the combined random measurement errors and the stochastic part of the true position walks of the source on the sky. The level of noise is drastically reduced by median binning of the sequence. The sorted sequence of tuples is divided into a large number of non-overlapping consecutive bins of 5468 points each. The median $\tau_i$ and the median $\delta_i$ are computed for each bin $i$ and for each coordinate separately. The resulting empirical functions $d_x(\tau_i)$ and $d_y(\tau_i)$ display hidden patterns in the time-correlated part of the stochastic process responsible for the measured trajectory of the source. For a source without any correlated signal in the measured trajectory, the ADAF for each coordinate is a flat function at zero, apart from the remaining white noise. A strictly cyclic motion characterized by repeated returns to the vicinity of the same position with a period $P$ produces a wavy pattern with ADAF crossing zero at equally spaced points separated by integer multiples of $P$ in $\tau$. 
\begin{figure}[h]%
\centering
\includegraphics[width=0.95\textwidth]{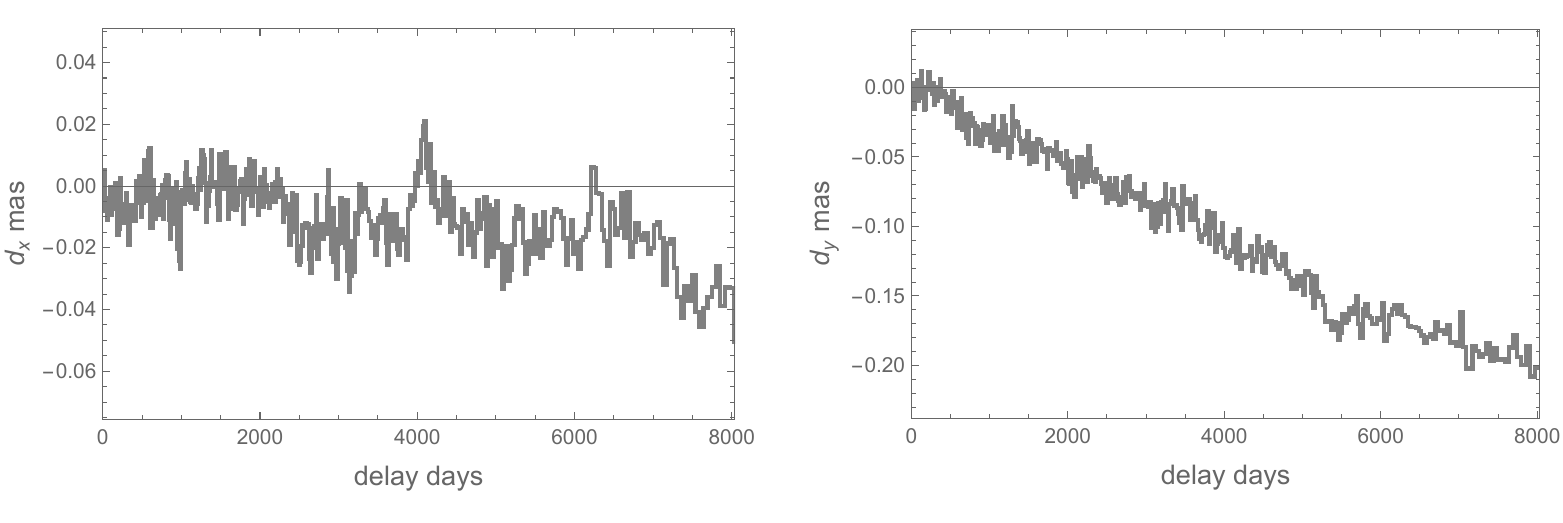}
\caption{Astrometric differential autoregressive functions of M87 in RA ($x$) and Dec ($y$) equatorial coordinates computed from 2269 24-hr VLBI sessions spanning 39 years.\\
The specific algorithm used for this computation is described in Sect. \ref{meth.sec}. The nearly linear trend in $d_y$ indicates a measurable proper motion in Dec.}\label{aaf.fig}
\end{figure}

The results of ADAF calculation are shown in Figs. \ref{aaf.fig} for each coordinate component. While the RA component shows a fairly flat dependence on $\tau$, which may be consistent with a purely white-noise pattern, the Dec component manifests a strong linear trend across the available range of $\tau$, which may be interpreted as a measurable proper motion. Our next task is to estimate this trend directly from the epoch positions.

\begin{figure}[ht]%
\centering
\includegraphics[width=0.45\textwidth]{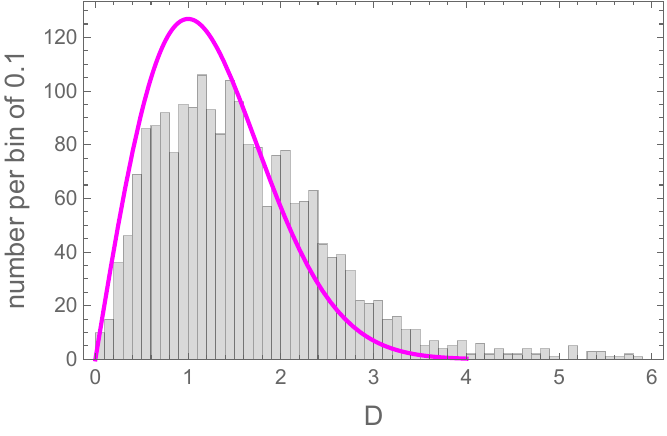}
\caption{Sample distribution of error-normalized astrometric offsets from the multi-decade mean position of the radio source M87. The magenta curve shows the expected Rayleigh[1] distribution if these bivariate values were Gaussian-distributed.}\label{D.fig}
\end{figure}

The model of astrometric trajectory used in this study includes a mean position $\{x_0,y_0\}=\{x(t_0),y(t_0)\}$ at a chosen mean epoch $t_0$ and linear trends in both coordinates with free coefficients called the proper motion components:
\begin{eqnarray}
    x_{\rm pm}(t)&=&x_0+\mu_x\,(t-t_0)+\nu_x \nonumber \\
    y_{\rm pm}(t)&=&y_0+\mu_y\,(t-t_0)+\nu_y.
\end{eqnarray}
The parallactic components, which are normally used in astrometry of stars, are not included because of the negligibly small values at intergalactic distances. The measurement noise components $\nu_x$, $\nu_y$ are modeled as a stationary random process. It is customary to solve these equations for $x_0$, $y_0$, $\mu_x$, $\mu_y$ using the least-squares method with weighting according to the given formal errors. As was shown in \citep{2024ApJS..274...28C}, the standard technique may provide perturbed results in the presence of strongly non-Gaussian components caused by the actual stochastic jitter and quasi-coherent displacements of the radio core on the sky. Fig. \ref{D.fig} shows that this is the case for M87, which is not a defining source in ICRF3 despite its brightness. The histogram is the sample distribution of the individual filtered, error-normalized position differences ``observed minus mean". The $D$-values were computed by Eq. 1 in \citep{2024ApJS..274...28C} using standard $2\times 2$ covariance matrices of the global astrometric solutions. The $D$-values are the 2D analogs of the unit weight errors for univariate statistics. The solid curve is the expected $\chi[2]$ distribution rescaled to the sample size. The measured offsets are not consistent with the assumed binormal distribution of individual measurements with the given covariances, mostly due to the presence of ``outliers" in the tail of the histogram. For example, 10\% of the sample are expected to have $D>2.146$, whereas we find 24\% of the measurements in excess of that value. Since the formal covariance involved in the least-squares fitting does not adequately capture the deviations from the assumed model, we opted for an unweighted robust solution of the condition equations using the 1-norm (L1) iterative optimization.

\begin{figure}[ht]%
\centering
\includegraphics[width=0.45\textwidth]{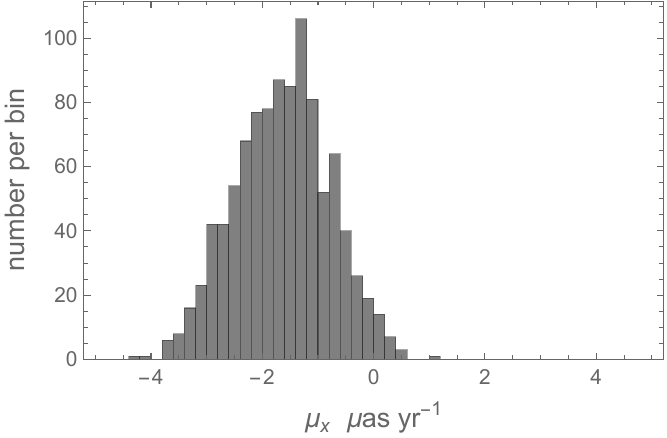}
\includegraphics[width=0.45\textwidth]{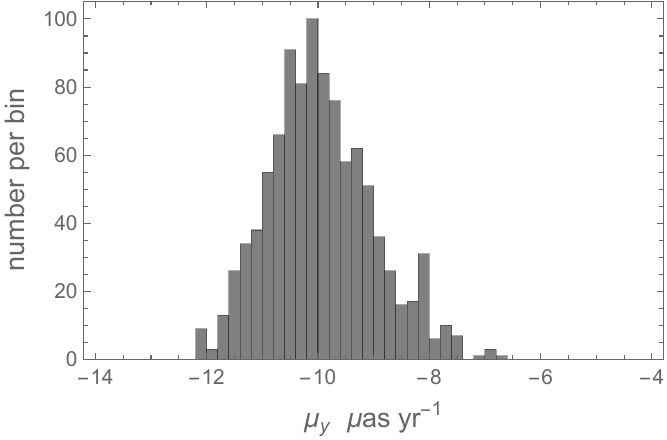}
\caption{Bootstrapped distributions of proper motion components of M87 estimated via Nelder-Meade 1-norm optimization.
Each computation (out of 1001) uses a randomly selected subsample of 1134 non-repeating data points.}\label{mu.fig}
\end{figure}

The optimization problem splits into two separate fits. For a given set of coordinate measurements $\{x_j,y_j\}$, the nonlinear objective function is
\eb 
\Psi(x_0,\mu_x)=\sum_{j=1}^n |x_j-x_{\rm pm}(t_j)|,
\label{psi.eq}
\ee 
and a similar function for $y_0$ and $\mu_y$ free parameters. Any standard nonlinear optimization method can be used for a sufficiently dense observational cadence.\footnote{We used the Nelder-Mead method, which is the default in Wolfram Mathematica} To estimate the uncertainty of the emerging estimates, we used the bootstrapping method, because the formal uncertainties are useless outside the least-squares optimization. The sample of 2269 single-epoch data points is randomly decimated to its half in 1001 trials. Each random selection (without replacement in the main sample) of 1134 data points is used to estimate the four parameters of the linear model \ref{psi.eq}. The resulting histograms of the proper motion components are depicted in Fig. \ref{mu.fig}. These bootstrapping sample distributions are further used to compute the robust estimates of statistical uncertainties $\sigma_{\mu x}$, $\sigma_{\mu y}$ as the difference of the 0.84 and 0.16-quantiles divided by $2\sqrt{2}$. Separately, sets of 1001 position angles and proper motion magnitudes are computed, and the corresponding uncertainties of these parameters are derived using the same quantile-based formula.

\section{Results}\label{res.sec}

\subsection{Distance functions and robust proper motion fits}
The ADAF analysis above serves as a model-independent diagnostic and nonparametric sanity check; the proper motion estimates reported below are derived independently from a direct fit to the unsmoothed single-epoch positions.
Fig. \ref{aaf.fig} shows the ADAF (Section \ref{meth.sec}) computed from the 2269 single-epoch measurements of M87 by the geodetic VLBI system between years 1986.3 and 2025.0. This source is one of the brightest and most frequently observed in ICRF3. By design, an ADAF originates from zero at zero time delay, and it is expected to diverge if there is a consistent trajectory of the source in the sky plane, because the positions separated by longer time spans are statistically more distant. We can see an obvious difference between the ADAF behavior in right ascension (RA, which is designated $x$ for brevity) and declination (Dec, or $y$) components. While the RA pairwise displacement mostly shows some small-amplitude undulations and noise, the Dec component is faithfully represented by a linear trend. The only type of motion consistent with a linear ADAF is the true linear motion with a constant rate. The ADAF analysis reveals the presence of a linear motion component, which is traditionally called proper motion in astrometry, in mostly the Dec coordinate. The magnitude of this motion is approximately 380 $\mu$as in 38 years.

We further apply the robust method of computing the proper motion from single-epoch coordinates, which is described in Sect. \ref{meth.sec}. Unlike the often used weighted least-squares optimization, which is only optimal for independent Gaussian errors with accurately known variances, the method produces more stable results in the presence of ``cosmic error" components \citep{2024ApJS..274...28C}. The non-weighted option is justified by the observation that the non-Gaussian cosmic error component dominates the observed dispersion of individual measurements of bright radio-loud AGNs, to which M87 belongs. The result from a nonlinear 1-norm optimization is $\mu=\{-1.65,-10.07\}$ $\mu$as yr$^{-1}$, i.e., the projected direction of motion is at an angle wider than $90\degr$ with respect to the jet. 

Since we chose the unweighted nonlinear fitting technique in favor of the often employed least-squares regression, the formal errors could not be used to estimate the statistical uncertainty of the fit. Instead, 1001 bootstrapping trials were performed. Each trial includes a random selection (without replacement) of a subset with half of the measurements, and an independent fit on the subset. The emerging proper motion components are used to compute the median values and the $1\sigma$-analog of the formal uncertainty, which is the difference of the 0.84- and 0.16-quantile divided by $2\sqrt 2$. The median proper motion values are estimated at $\{-1.60,-10.03\}$ $\mu$as yr$^{-1}$, which is close to the overall sample fit. The estimated uncertainties are $\{0.63,0.65\}$ $\mu$as yr$^{-1}$. Separately, we directly estimate the magnitude of proper motion $|\mu|=10.19\pm 0.64$ $\mu$as yr$^{-1}$ and position angle $\theta=189.2^\circ \pm 3.5^\circ$ from the bootstrap results. 
We note that the RA and Dec components are projections of the single 2D fit and are reported for reference; the statistical significance of the detection rests on the proper motion magnitude and its uncertainty, estimated directly from the bootstrap distribution.
\begin{figure}[h]%
\centering
\includegraphics[width=0.9\textwidth]{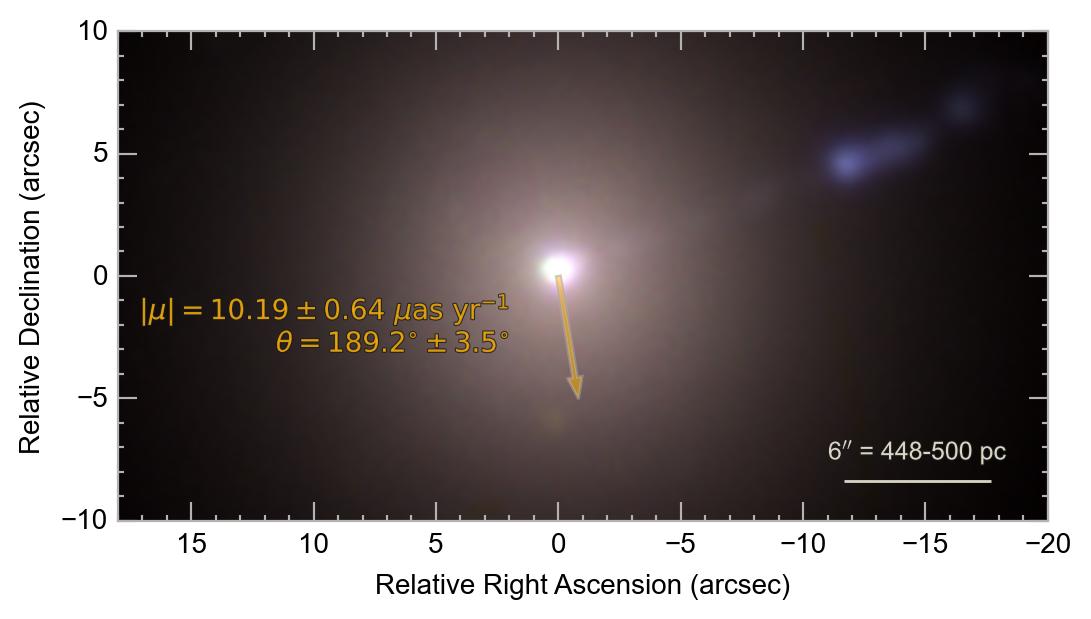}
\caption{Composite image of M87 with the proper motion vector derived in this study. \\
The component $g,z,y$ filter images were extracted from the Pan-STARRS imaging facility available at \url{https://ps1images.stsci.edu/cgi-bin/ps1cutouts}. North is up, east is to the left, and the angular scale bar also describes the corresponding physical scales for the range of estimated distances to M87. The jet is seen at position angle 288$^\circ$. 
The arrow shows the direction (E from N) of our calculated proper motion vector, and the length displayed here corresponds to the apparent location it would occupy after 0.5 Myr at a constant rate.
}
\label{map.fig}
\end{figure}

Fig. \ref{map.fig} displays a composite optical image of M87 with a yellow arrow indicating the direction of the proper motion vector determined in this study. This direction is completely inconsistent with the axis of the relativistic jet at position angle $288^\circ$. This suggests that the apparent long-term motion of the AGN is not related to structural changes in the inner jet, which are sometimes invoked to explain the trajectories of ICRF quasars \citep{2024ApJ...977L..14M}. M87 belongs to the group of radio-loud AGNs with misaligned jet structures and apparent motion \citep{2011AJ....141..178M}. High-resolution imaging at shorter radio wavelengths revealed a slowly moving or stationary helical inner jet structure pointing in the same direction as the optical jet \citep{2018ApJ...855..128W}. To test the possible effect of nonzero correlations between the RA and Decl components in the VLBI measurements, we reproduced the proper motion fit in a local coordinate frame rotated counterclockwise through $198\degr$, thus aligning the $X'$ axis with the jet direction at PA$=288\degr$. The components of proper motion in this frame are $\mu_{x'}=-1.01\pm 0.81$ and $\mu_{y'}=10.23\pm 0.79$ $\mu$as yr$^{-1}$ comfirming its nearly orthogonal orientation with respect to the jet. Surprisingly, the proper motion direction is close to the position angle of the clearly visible diffuse source separated by $6.09$ arcsec from the nucleus in position angle $178^\circ$, which is associated with source 3907709443748705280 in Gaia DR3 \citep{2023A&A...674A...1G}. 
This source was identified as a globular cluster in M87 by \cite{2009ApJ...703...42P}, with coordinates ($\alpha, \delta) = (187.7060037, 12.3894173)^\circ$.

\subsection{Robustness of the Proper Motion Estimate}
\label{robustness.sec}

The astrometric position time series analyzed in this work is a product of a global geodetic VLBI solution in which station positions, station velocities, Earth orientation parameters, and source coordinates are estimated simultaneously in a single least-squares optimization. Secular effects in the terrestrial network — such as baseline length changes, station velocity errors, or site-specific systematics — are explicitly modeled and absorbed into the TRF solution, and do not propagate as unmodeled biases into the source position estimates. The celestial reference frame is realized by imposing a no-net-rotation constraint on the ensemble of ICRF3 defining sources; M87 is estimated independently as a non-defining arc parameter and is therefore unconstrained in its positional evolution. Any residual instability in the frame realization — for instance, due to the evolving composition of the defining source set over time — would manifest as a coherent low-order signal (dipolar or quadrupolar) across many sources, rather than as an isolated linear drift in a single object. Any low-frequency variations of the geodetic system of this provenance would be directly noticed in the Earth orientation parameters, which in fact appear to be constrained to $<1$ $\mu$as.

Several features of the analysis provide additional confidence in the result. The proper motion is estimated via a 2D L1-norm optimization on the complete set of 2269 single-epoch positions, with uncertainties derived from 1001 bootstrap trials. This approach is insensitive to the outlier population that dominates the tails of the error distribution (Figure \ref{D.fig}). The ADAF diagnostic (Figure \ref{aaf.fig}) provides an independent, nonparametric confirmation of a linear trend in declination across the full range of time delays. The position time series exhibits epoch-to-epoch scatter that varies with the size and geometry of the observing network; this heteroscedasticity motivates our use of the unweighted L1 method, which is robust to the non-Gaussian outlier population documented for bright radio-loud AGNs \citep{2024ApJS..274...28C}. We emphasize that the smoothed time series presented in Figures \ref{times.fig} and \ref{VLBI_timeseries.fig} serves only as a visualization aid; all quantitative analysis is performed on the unsmoothed per-session position estimates. The short-term coherent astrometric variations visible in the smoothed data, on timescales of months to years, are attributed to a combination of source structure variability, network geometry effects, and seasonal or loading-related signals in the geodetic solution. Using the robust periodogram method from \citep{2024PASP..136e4503M}, separate amplitude periodograms were computed for position offset components of M87 between 8 d and 10 yr, which were found to be consistent with white noise. No periodic signals were detected within the corresponding range of frequencies exceeding the 99\% confidence threshold.

\section{Conclusions}

The range of distance estimates for M87 in the literature is from 15.4 Mpc \citep{2019AstBu..74..257T} to 17.2 Mpc \citep{2007ApJ...655..144M} with a typical quoted uncertainty ($1\sigma$) of about 0.6 Mpc. The heliocentric radial velocity of the galaxy is 1284$\pm 5$ km s$^{-1}$ \citep{2011MNRAS.413..813C}. From the spectroscopic redshift measurement ($z=0.0042$), the expected radial velocity is $\sim 1260$ km s$^{-1}$, which is reasonably close. Assuming a Hubble parameter value of $H_0=75$ km s$^{-1}$ Mpc$^{-1}$, the cosmological expansion component comes up to between 1155 and 1290 km s$^{-1}$. The peculiar radial velocity of M87 is then approximately between 0 and $+130$ km s$^{-1}$, i.e., the galaxy is at rest or slowly receding from us. The barycentric tangential velocity estimated in this study is a much higher 743--830 km s$^{-1}$ for the range of distances. Thus, the relative motion of M87 with respect to the Solar system barycenter is mostly tangential to the radial direction. It is practical at this point to transfer the estimated proper motion from the equatorial to the Galactic coordinate frame, which yields $\{\mu_l,\mu_b\}=\{1.45,-10.08\}$ $\mu$as yr$^{-1}$. We can now compute the 3D velocity vector of M87 in the Galactic coordinate system relative to the Solar system barycenter. The largest uncertainty comes from the imprecisely known radial velocity with respect to the Hubble flow and the estimated distance. Assuming a radial velocity of 100 km s$^{-1}$, which is close to the median of available estimates, the absolute barycentric velocity of M87 is between 750 (for distance $D=15.4$ Mpc) and 836 km s$^{-1}$ (for $D=17.2$ Mpc).

The Solar system observer moves with a velocity of $369.82\pm 0.11$ km s$^{-1}$ toward $\{l,b\}=\{264.021^\circ,48.243^\circ \}$ in the inertial cosmic frame, which is at rest with respect to the CMB \citep{2020A&A...641A...6P}. This velocity vector comprises the orbital motion of the Solar system in the Galaxy and the peculiar velocity of the Milky Way with respect to the CMB. The corresponding 3D velocity vector of the observer in the Galactic frame is $[-26,\,-245,\,+276]^T$ km s$^{-1}$. This vector should be added to the estimated barycentric velocity of M87 $[+278,\,-689,\,-100]^T$ km s$^{-1}$ (for $D=15.4$ Mpc) to obtain the velocity of M87 in the cosmic frame. The result is a vector with a magnitude of 983 km s$^{-1}$ (or 1062 km s$^{-1}$ for $D=17.2$) in the direction of approximately $(l,b)=(285^\circ,9^\circ)$. The angle between our line of sight and this velocity vector is $65^\circ$. The median speed corresponding to $D=16.1$ Mpc is $1037\pm 30$ km s$^{-1}$.

Our determination of the peculiar velocity of M87 can be used for testing the theoretical prediction of the local kinematics or as a constraint on the developed $\Lambda$CDM models. In the supergalactic coordinate system, the estimated barycentric peculiar velocity vector is between $[-671,-65,-328]^T$ and $[-747,-84,-366]^T$ km s$^{-1}$ for the range of distances from 15.4 to 17.2 Mpc. The angular supergalactic coordinates of the radio core are SGL $=102.881^\circ$ and SGB $=-2.348^\circ$. This implies that the motion of M87 in the supergalactic $XY$ plane is nearly orthogonal to the line of sight toward negative SGX values. The angle between the line of sight and the Galactocentric mean velocity of M87 is estimated at $86\degr$.

The relative 3D velocity of M87 with respect to the center of the Milky Way provides valuable information about the relative motion of galaxies and clusters within the Local Supecluster (LSC). Using the latest estimate of the astrometric proper motion of the central radio source Sgr A obtained with the VLBI facility in the $K$-band \citep{2023AJ....165...49G}, $\{\mu_{\alpha*},\mu_\delta\}=\{-3.128,-5.584\}\pm\{0.042, 0.075\}$ mas yr$^{-1}$, the corresponding velocity components in the Galactic coordinates are approximately $\{-248, -9\}$ km s$^{-1}$ assuming a solar radius of 8.18 kpc from \citep{2019A&A...625L..10G}. The radial component of the solar barycenter motion is not known from the astrometric measurements, but we can assume that it is equal to the first coordinate component of the Solar system peculiar velocity with respect to the Local Standard of Rest, which is equal to 12.24 km s$^{-1}$ from \citep{10.1111/j.1365-2966.2010.16253.x}. The resulting Galactic velocity vector of the solar barycenter is $[12,\,248,\,9]^T$ km s$^{-1}$. This vector should be added to the previously estimated barycentric velocity of M87 in this paper, resulting in a range of estimates between $[+290,\,-441,\,-91]^T$ and $[+322,\,-519,\,-114]^T$ km s$^{-1}$, depending on the assumed distance to M87. Hence, the estimated velocity of M87 relatively to the Milky Way (with the cosmic expansion component taken out) ranges between 536 and 621 km s$^{-1}$, and its direction is almost perfectly transverse, the component along the line connecting the two galaxies being just 45 km s$^{-1}$. The latter value is fairly consistent with the previously estimated peculiar radial velocity 100 km s$^{-1}$.

For a more exact comparison with cosmological and numerical predictions of the local velocity field in the framework of the $\Lambda$CDM model, the Galactocentric vector should be converted to the supergalactic coordinates, which yields $[-550, -86, -156]^T$ km s$^{-1}$ for a distance $D=16.1$ Mpc. 
Is this consistent with our understanding of the local Universe's kinematics? Based on the data collected in the CosmicFlows-3 dataset \citep{2016AJ....152...50T}, Shaya et al. \citep{2017ApJ...850..207S} reconstructed a 3D history of bulk motion inside the LSC. The Virgo cluster, which is the largest part of the LSC near its apparent center, is almost at rest with respect to the center of mass of the  LSC according to their results. The entire supercluster, on the other hand, is involved in a large-scale flow toward the Great Attractor, which is directed predominantly toward negative SGX \citep{2003ApJ...596...19K}.
Our result appears to be consistent with the direction to the Great Attractor, which has approximate supergalactic coordinates [SGX,SGY,SGZ]$_{\rm GA}=[-0.972,0.175,-0.159]^T\times 58$ Mpc. Indeed, the angle between the Galactocentric velocity of M87 and the direction toward the Great Attractor is just $20^{\circ}$. However, this interpretation of our results implies that the Milky Way is not involved in the bulk motion of the LSC toward the Great Attractor, while M87 is.

As the main caveat, our conclusions are implicitly based on the supposition that the apparent motion of the AGN at the center of M87 represents the motion of its center of mass (and a similar assumption is made about the Milky Way and the Sgr A radio-emitting AGN). Physically feasible mechanisms have been proposed violating this condition. One of the most frequently considered possibilities is an offset of the central black hole (BH) from the center of the gravitational potential of the host galaxy caused by a coalescence event in a supermassive BH binary. Such an event is known to produce recoil velocities of several hundreds km s$^{-1}$ \citep{2007arXiv0710.1338P}. The merged BH and the associated AGN would continue to travel in loose orbital loops around the true center of mass until the damping force from the interacting medium (gas and stars) and relativistic outflows remove the excess kinetic momentum and the BH settles back at the center \citep{2008ApJ...678..780G}. For a recent ejection event, neglecting the damping forces and assuming a uniform density $\rho$ of gravitating mass in the core, the motion of the BH is a harmonic oscillator with a period $\sqrt{3\pi/(G\,\rho)}$. This period equals 45.8 Kyr for a core density $\rho=10^6 M_\odot$ per pc$^3$. The maximum separation amounts to 7.4 pc for an initial kick velocity of 1000 km s$^{-1}$ in the galaxy's rest frame. The measured velocity in this case is the sum of the proper velocity of M87 in space and the relative velocity of the BH within the galaxy. A projected displacement of the point-like source (presumably, AGN) from the centers of optical isophotes by 6.8 pc was reported in the literature \citep{2010ApJ...717L...6B}, but its position angle is consistent with the direction of the jet, and a more careful analysis revealed that this displacement may only be a technical feature of the isophotal technique on extended images with time-variable structures \citep{2018MNRAS.480.4099L}. On the other hand, the high-resolution images of the central region surrounding the BH obtained with the Event Horizon Telescope revealed an asymmetric crescent structure (instead of the expected ring) with the southern segment showing elevated brightness \citep{2020A&A...634A..38N}. Additional observations confirmed the stable radius of the crescent (43.3 $\mu$as) and revealed possible variations in the brightness structure with time \citep{2024A&A...681A..79E}, which may be sensitive to the image priors used for the 2D image reconstruction from 1D VLBI scans \citep{2024ApJ...975..201F}, but confirming the enhanced brightness on the southern limb. Speculatively, a relative motion of the BH in the galactic medium may generate an asymmetry in the ring structure.

\section*{Supplementary information}
An animation of the VLBI source position time series presented in Fig.~\ref{VLBI_timeseries.fig} is given here. 
STR120d\_2Dpolar\_3C274\_fromMJD49600.jpg

\section*{Acknowledgments}

This work supports USNO's ongoing research into the celestial reference frame and geodesy. 
The National Radio Astronomy Observatory is a facility of the National Science Foundation operated under cooperative agreement by Associated Universities, Inc. 
The authors acknowledge use of the Very Long Baseline Array under the U.S.\ Naval Observatory’s time allocation.  
The authors have used the IVS data archive maintained by the International VLBI Service for Geodesy and Astronomy, \url{https://ivscc.gsfc.nasa.gov/products-data/index.html}. The authors thank Nathan Secrest for his valuable insights and stimulating discussions regarding cosmological implications of the results and updating the value of the Galactic aberration effect.

\bibliography{main}{}
\bibliographystyle{aasjournalv7}
\end{document}